\documentclass{kluwer}

\usepackage{graphicx}

\begin{document}
\begin{article}


\begin{opening}

\title{Analyze of the star formation modeling algorithm in SPH code}

\author{Peter \surname{Berczik}}

\runningauthor{Peter Berczik}
\runningtitle{Analyze of the star formation modeling algorithm in SPH code}

\institute{Main Astronomical Observatory of Ukrainian National Academy of Sciences, \\
UA-03680, Golosiiv, Kiev-127, Ukraine, e-mail: {\tt berczik@mao.kiev.ua}}

\date{November 17, 2000}


\begin{abstract} The chemical and photometric evolution of star forming
disk galaxies is investigated. Numerical simulations of the complex
gasdynamical flows are based on our own coding of the Chemo - Dynamical
Smoothed Particle Hydrodynamical (CD - SPH) approach, which incorporates the
effects of star formation. As a first application, the model is used to
describe the chemical and photometric evolution of a disk galaxy like the Milky
Way. \end{abstract}



\end{opening}


\section{Introduction}

Galaxy formation is a highly complex subject requiring many different
approaches of investigation. Recent advances in computer technology and
numerical methods have allowed detailed modeling of baryonic matter dynamics in
a universe dominated by collisionless dark matter and, therefore, the detailed
gravitational and hydrodynamical description of galaxy formation and evolution.
The most sophisticated models include radiative processes, star formation and
supernova feedback, e.g. \cite{K92,SM94,FB95}.

The results of numerical simulations are fundamentally affected by the star
formation algorithm incorporated into modeling techniques. Yet star formation
and related processes are still not well understood on either small or large
spatial scales. Therefore the star formation algorithm by which gas is
converted into stars can only be based on simple theoretical assumptions or on
empirical observations of nearby galaxies.

Among the numerous methods developed for modeling complex three dimensional
hydrodynamical phenomena, Smoothed Particle Hydrodynamics (SPH) is one of the
most popular \cite{M92}. Its Lagrangian nature allows easy combination with
fast N - body algorithms, making possible the simultaneous description of
complex gas-stellar dynamical systems \cite{FB95}. As an example of such a
combination, TREE - SPH code \cite{HK89,NW93} was successfully applied to
the detailed modeling of disk galaxy mergers \cite{MH96} and of galaxy
formation and evolution \cite{K92}. A second good example is an GRAPE - SPH
code \cite{SM94,SM95} which was successfully used to model the evolution of
disk galaxy structure and kinematics.


\section{The model}

The hydrodynamical simulations are based on our own coding of the Chemo -
Dynamical Smoothed Particle Hydrodynamics (CD - SPH) approach, including
feedback through star formation (SF). The dynamics of the "star" component is
treated in the framework of a standard N - body approach. Thus, the galaxy
consists of "gas" and "star" particles. For a detailed description of the CD -
SPH code (the star formation algorithm, the SNII, SNIa and PN production, the
chemical enrichment and the initial conditions) the reader is referred to
\cite{BerK96,Ber99,Ber2000}. Here we briefly describe the basic features of our
algorithm.

\subsection{Star formation algorithm}

It is well known that SF regions are associated with giant molecular complexes,
especially with regions that are approaching dynamical instability. The overall
picture of star formation seems to be understood, but the detailed physics of
star formation and accompanying processes, on either small or large scales,
remains sketchy \cite{L69,S87}.

All the above stated as well as computer constraint cause the using of
simplified numerical algorithms of description of conversion of the
gaseous material into stars, which are based on simple theoretical
assumptions and/or on results of observations of nearby galaxies.

We modify the "standard" SPH SF algorithm \cite{K92,SM94,SM95}, taking into
account the presence of chaotic motion in the gaseous environment and the time
lag between the initial development of suitable conditions for SF, and SF
itself.

Inside a "gas" particle, the SF can start if the absolute value of the "gas"
particles gravitational energy exceeds the sum of its thermal energy and its
energy of chaotic motions:
\begin{equation}
  \mid E_i^{gr} \mid > E_i^{th} + E_i^{ch}.
\end{equation}
Gravitational and thermal energies and the energy of random motions for
the "gas" particle $ i $ in model simulation are defined as:
\begin{eqnarray}
  \left\{
  \begin{array}{lllll}
  E_i^{gr} = - \frac{3}{5} \cdot G \cdot m_i^2/h_i,    \\
                                                       \\
  E_i^{th} = \frac{3}{2} \cdot m_i \cdot c_i^2,        \\
                                                       \\
  E_i^{ch} = \frac{1}{2} \cdot m_i \cdot \Delta v_i^2, \\
  \end{array}
  \right.
\end{eqnarray}
where $ c_i = \sqrt{\Re \cdot T_i / \mu} $ is the isothermal sound
speed of particle $ i $. We set $ \mu = 1.3 $ and define the random or
"turbulent" square velocities near particle $ i $ as:
\begin{equation}
  \Delta v_i^2 = \sum_{j=1}^{N_{B}} m_j \cdot ({\bf v}_j - {\bf v}_c)^2 /
                 \sum_{j=1}^{N_{B}} m_j,
\end{equation}
where:
\begin{equation}
  {\bf v}_c = \sum_{j=1}^{N_{B}} m_j \cdot {\bf v}_j /
              \sum_{j=1}^{N_{B}} m_j.
\end{equation}
For practical reasons, it is useful to define a critical temperature
for SF onset in particle $ i $ as:
\begin{equation}
  T^{crit}_i = \frac{\mu}{3 \Re} \cdot
               ( \frac{8}{5} \cdot \pi \cdot G \cdot \rho_i \cdot h_i^2
               - \Delta v_i^2 ).
\end{equation}
Then, if the temperature of the "gas" particle $ i $, drops below the
critical one, SF can proceed.
\begin{equation}
  T_i < T^{crit}_i.
\end{equation}
The chosen "gas" particle produces stars only if the above condition holds over
the time interval exceeding its free - fall time:
\begin{equation}
  t_{ff} = \sqrt { \frac{3 \cdot \pi}{32 \cdot G \cdot \rho} }.
\end{equation}
We also check that the "gas" particles remain cool, i.e. $ t_{cool} < t_{ff} $.
We rewrite these conditions following \cite{NW93}:
\begin{equation}
  \rho_i > \rho_{crit}.
\end{equation}
We set the value of $\rho_{crit} = 0.03$ cm$^{-3}$.

When the collapsing particle $ i $ is defined, we create the new "star"
particle with mass $ m^{star} $ and update the "gas" particle $ m_i $ using
these simple equations:
\begin{eqnarray}
  \left\{
  \begin{array}{lll}
  m^{star} = \epsilon  \cdot m_i,  \\
                                  \\
  m_i = (1 - \epsilon) \cdot m_i.  \\
  \end{array}
  \right.
\end{eqnarray}
In the Galaxy, on the scale of giant molecular clouds, the typical values for
SF efficiency are in the range $ \epsilon \approx 0.01 \div 0.4 $
\cite{DIL82,WL83}.

We did not fix this value but rather also derived $ \epsilon $ from the
"energetic" condition:
\begin{equation}
  \epsilon = 1 - \frac{E_i^{th} + E_i^{ch}}{\mid E_i^{gr} \mid}.
\end{equation}
At the moment of birth, the positions and velocities of new "star" particles
are set equal to those of parent "gas" particles. Thereafter these "star"
particles interact with other "gas" and "star" or "dark matter" particles only
by gravity.

\subsection{Thermal SNII feed--back}

For the thermal budget of the ISM, SNIIs play the  main role. Following to
\cite{K92,FB95}, we assume that the energy from the explosion is converted
totally to thermal energy. The total energy released by SNII explosions ($
10^{44} \; J $ per SNII) within "star" particles is calculated at each time
step and distributed uniformly between the surrounding "gas" particles
\cite{RVN96}.

\subsection{Chemical enrichment of gas}

In our SF scheme, every new "star" particle represents a separate,
gravitationally bound, star formation macro region (like a globular cluster).
The "star" particle has its own time of birth $ t_{begSF} $ which is set equal
to the moment the particle is formed. After the formation, these particles
return the chemically enriched gas into surrounding "gas" particles due to
SNII, SNIa and PN events.

We concentrate our treatment only on the production of $^{16}$O and $^{56}$Fe,
yet attempt to describe the full galactic time evolution of these elements,
from the beginning up to present time (i.e. $ t_{evol} \approx 15.0 $ Gyr).

\subsection{Photometric evolution of star component}

The code also includes the photometric evolution of each "star" particle, based
on the idea of the Single Stellar Population (SSP) \cite{BCF94,TCBF96}.

At each time - step, absolute magnitudes: M$_U$, M$_B$, M$_V$, M$_R$, M$_I$,
M$_K$, M$_M$ and M$_{bol}$ are defined separately for each "star" particle. The
SSP integrated colours (UBVRIKM) are taken from \inlinecite{TCBF96}. The spectro -
photometric evolution of the overall ensemble of "star" particles forms the
Spectral Energy Distribution (SED) of the galaxy.

We do not model the energy distribution in spectral lines nor the scattered
light by dust. However according to \inlinecite{TCBF96} our approximation is
reasonable, especially in the UBV spectral brand.

\subsection{Initial conditions}

As the initial model (relevant for CDM - scenario) we took a constant--density
homogeneous gaseous triaxial configuration ($ M_{gas} = 10^{11} \; M_\odot $)
within the rigid Plummer--type dark matter halo ($ M_{halo} = 10^{12} \;
M_\odot $). We set the scale length of dark matter halo: $ b_{halo} = 25 $ kpc.
These values of $ M_{halo} $ and $ b_{halo} $ are typical for dark matter halo
in disk galaxies \cite{NFW96,NFW97,B95}. We set $ A = 100 $ kpc, $ B = 75 $ kpc
and $ C = 50 $ kpc for semiaxes of system. Such triaxial configurations are
reported in cosmological simulations of the dark matter halo formation
\cite{EL95,FWDE88,WQSZ92}.

The gas was assumed to be involved in the Hubble flow ($ H_{0} = 65 $ km/s/Mpc)
and the solid - body rotation around $ z $ - axis. The spin parameter in our
numerical simulations is $ \lambda \approx 0.08 $, defined in \inlinecite{P69}
as:
\begin{equation}
  \lambda = \frac{\mid {\bf L}_0 \mid \cdot \sqrt{\mid E_0^{gr} \mid}}
              {G \cdot (M_{gas}+M_{halo})^{5/2}},
\end{equation}
$ {\bf L}_0 $ is the total initial angular momentum and $ E_0^{gr} $ is
the total initial gravitational energy of a protogalaxy. It is to be
noted that for a system in which angular momentum is acquired through
the tidal torque of the surrounding matter, the standard spin parameter
does not exceed $ \lambda \approx 0.11 $ \cite{SB95}. Moreover, its
typical values range between $ \lambda \approx 0.07^{+0.04}_{-0.05} $,
e.g. $ 0.02 \le \lambda \le 0.11 $.


\section{Results and discussion}

Our SPH calculations were carried out for the three different number of "gas"
particles $N_{gas} = 2109; \; 4109; \; 11513$. As a "base" model we consider the
case with $N_{gas} = 2109$. For this case we calculate the full evolution of
galactic system (i.e. $t_{evol} \approx 15.0$ Gyr). According to
\inlinecite{NW93} and \inlinecite{RVN96}, such a number seems adequate for a
qualitatively correct description of the systems behaviour. Even this small a
number of "gas" particles produces $ N_{star} = 31631 $ "star" particles at the
end of the calculation. The other two cases: $N_{gas} = 4109$ and $11513$ we
need for the more extensive study of our SF scheme. In the case of $N_{gas} =
4109$ we calculate the evolution up to $t_{evol} \approx 5.0$ Gyr and in the
case $N_{gas} = 11513$ up to $t_{evol} \approx 2.5$ Gyr. In all this two cases
within these time scale the total number of particles access the $N_{max} =
65535$ and after this we, stop any SF activity.

The more extended discussion of the dynamical and chemical data can
be found in the papers \inlinecite{Ber99} and \inlinecite{Ber2000}. Here we mainly
concentrate our attention to the extensive check of our SF scheme and the
photometric behaviour of our "base" galaxy model.

\begin{figure}
\tabcapfont
\centerline{%
\begin{tabular}{c@{\hspace{0.1in}}c}
\includegraphics[width=2.5in]{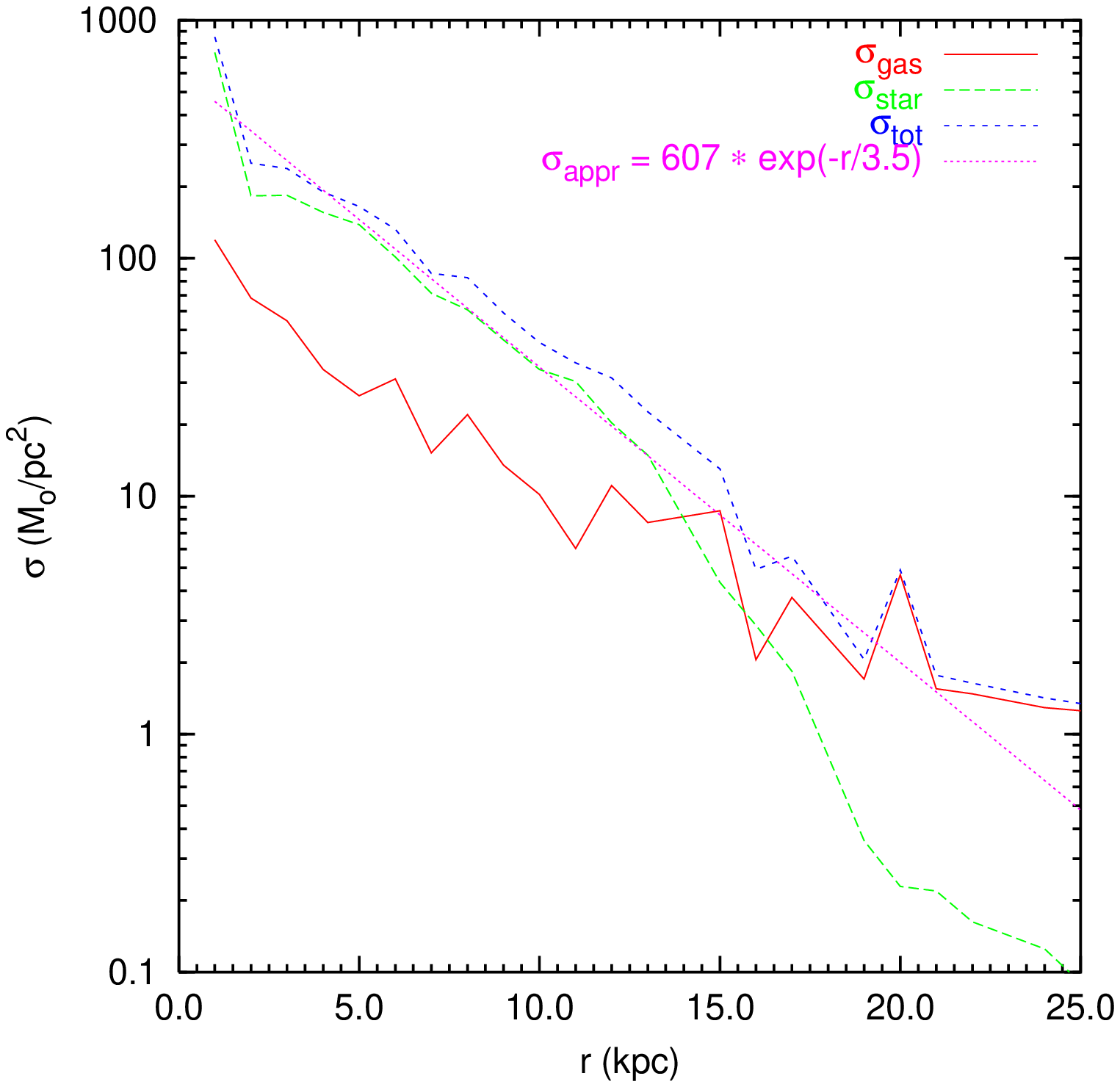} &
\includegraphics[width=2.5in]{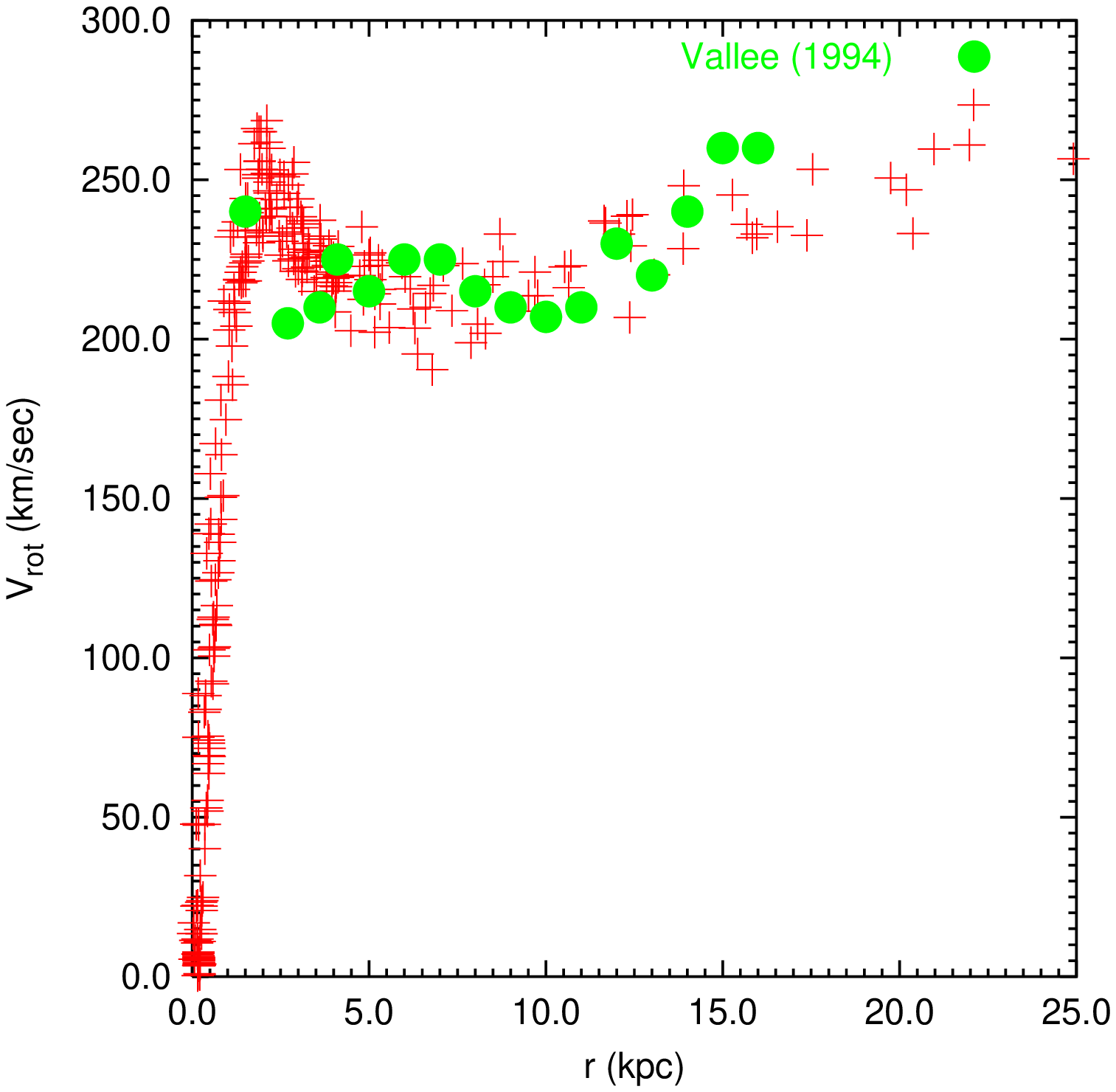} \\
a.~$\sigma$ distribution. & b.~$V_{rot}$ distribution.
\end{tabular}}
\caption{Column density \& rotational velocity distribution in the final step for the "basic" model.}
\label{sigma&vrot}
\end{figure}

In Figure~\ref{sigma&vrot} we present the column density distribution -(a) and
the rotational velocity distribution - (b) for our "base" galaxy model at the last
time step. The total density distribution ($\sigma_{tot} = \sigma_{gas} +
\sigma_{star}$) is well approximated by an exponential disk with radial scale
length $\sim 3.5$ kpc. The rotational velocity ($V_{rot}$) distribution also
well coincides with the data for our own galaxy \cite{Va94}.

\begin{figure}
\tabcapfont
\centerline{%
\begin{tabular}{c@{\hspace{0.1in}}c}
\includegraphics[width=2.5in]{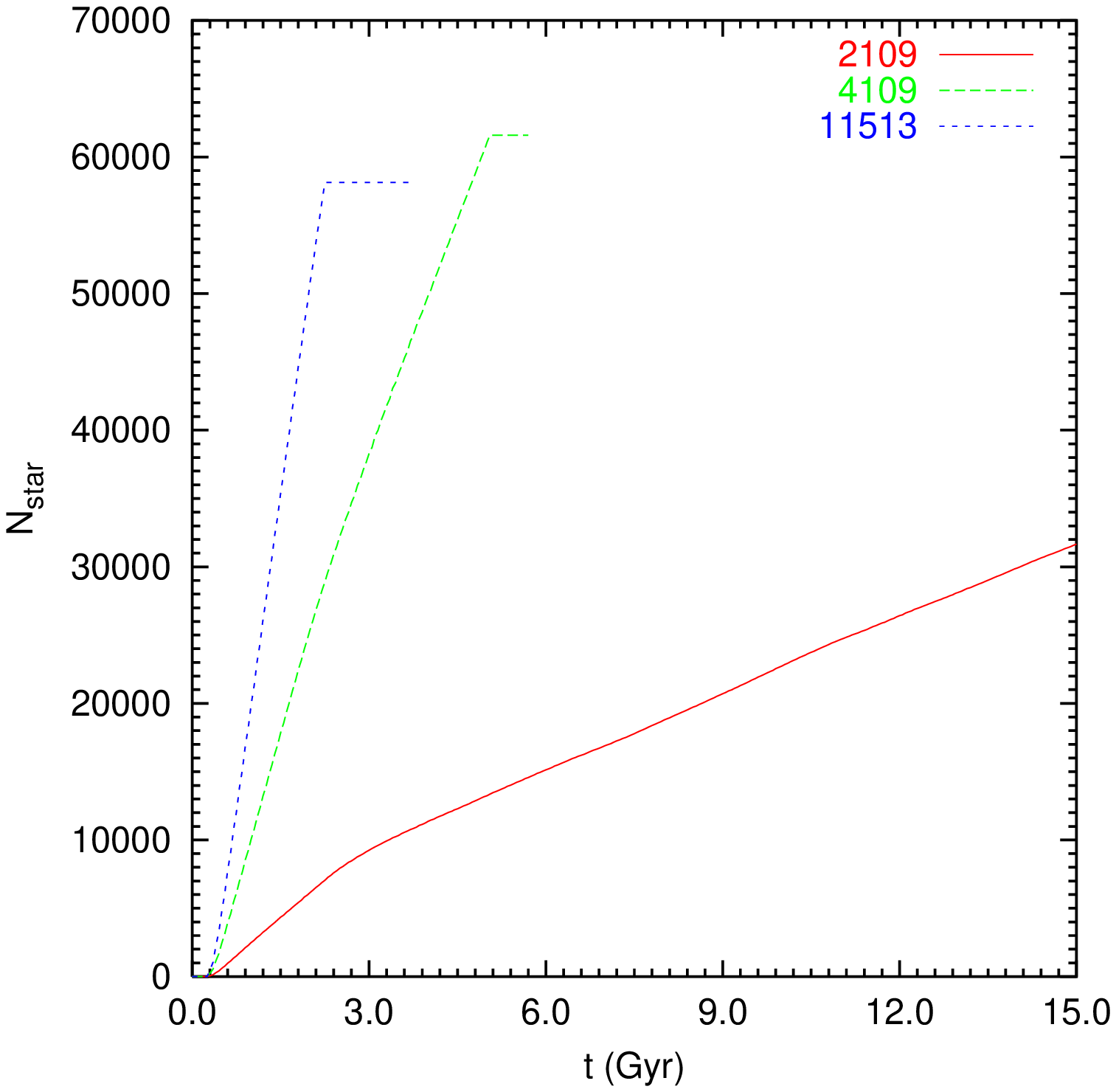} &
\includegraphics[width=2.5in]{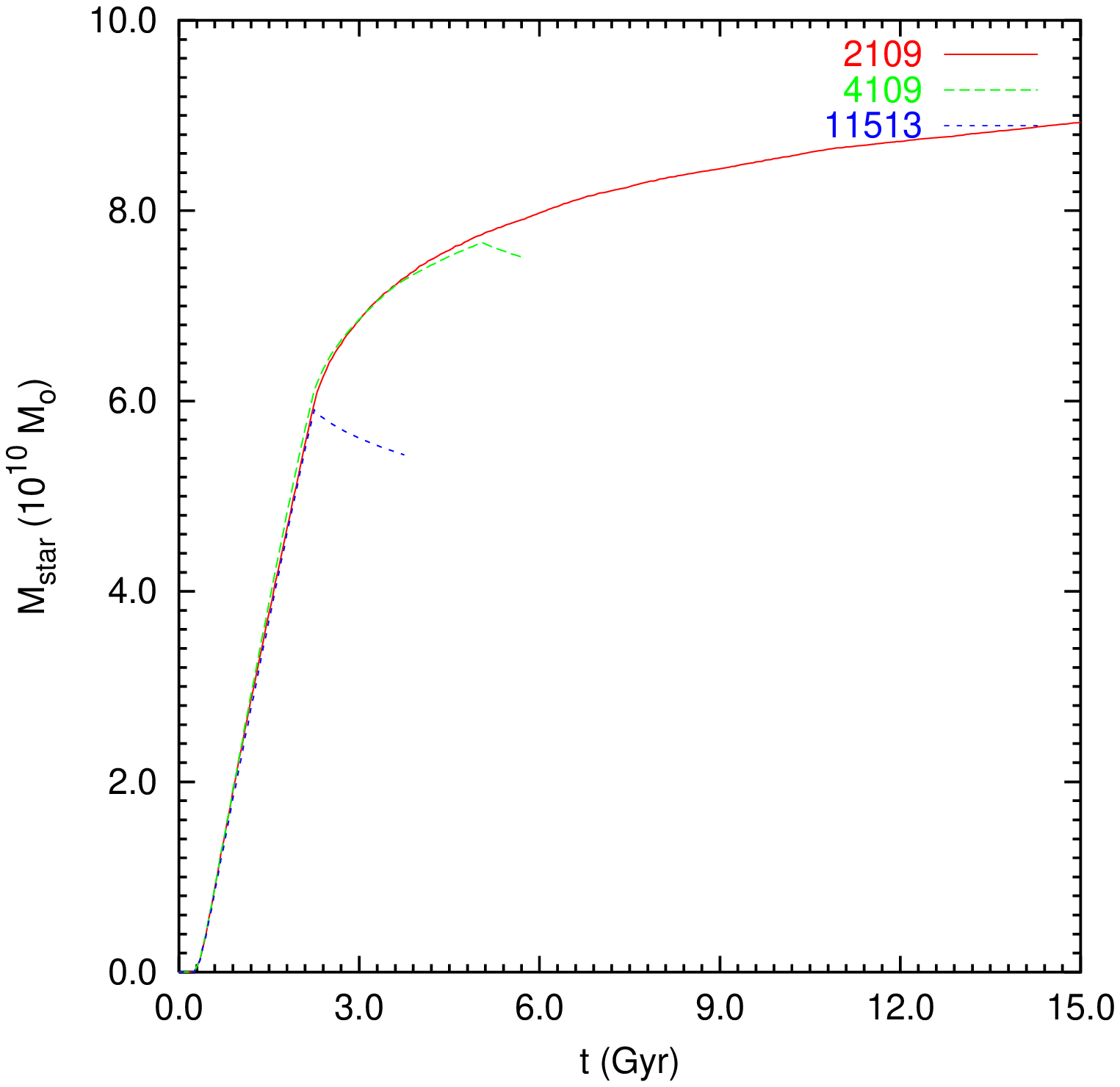} \\
a.~N$_{star}$ evolution. & b.~M$_{star}$ evolution.
\end{tabular}}
\caption{N$_{star}$ \& M$_{star}$ evolution of the model galaxy with different "gas" particle number.}
\label{nstar&mstar}
\end{figure}

In Figure~\ref{nstar&mstar}a we present the time evolution of "star" particles
number $N_{star}$. In Figure~\ref{nstar&mstar}b we present the time evolution
of the total mass inside "star" particles $M_{star}$. The number of "star"
particle is very different in all three cases, but the total mass inside the
"star" component is very similar. Some differences are found only after the SF
artificially stops in the models with high "gas" particle numbers. After stop
of SF the mass inside "star" component is reduced, according to returning the
mass from "star" to surrounding "gas" particles due to SNII, SNIa and PN
events.

\begin{figure}
\centerline{\includegraphics[width=3.5in]{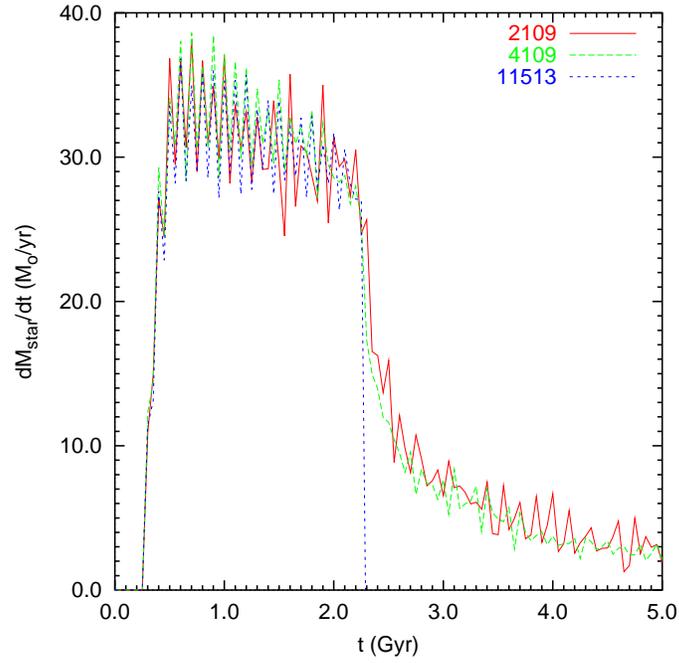}}
\caption{dM$_{star}/dt$ evolution of the model galaxy with different "gas" particle number.}
\label{sfh}
\end{figure}

\begin{figure}
\centerline{\includegraphics[width=3.5in]{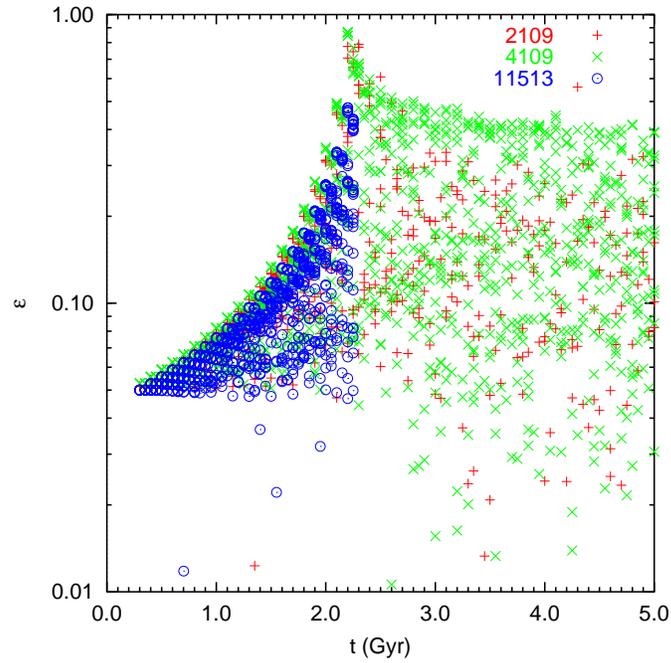}}
\caption{$\epsilon$ evolution of the model galaxy with different "gas" particle number.}
\label{sfe}
\end{figure}

The star formation history (SFH - $dM_{star}/dt$) for all three models is
presented in Figure~\ref{sfh}. In this figure we clearly see that the SFH
practically does not depend on the "gas" or "star" particle number. The star
formation efficiency (SFE - $\epsilon$) in each act of SF is presented in the
Figure~\ref{sfe}. The averaged SFE is about $\sim 10 \%$. However we have a
wide spread from some percent up to the $50 \%$ at the late stage of evolution.
In all three models we have a similar SFE trend.

\begin{figure}
\centerline{\includegraphics[width=3.5in]{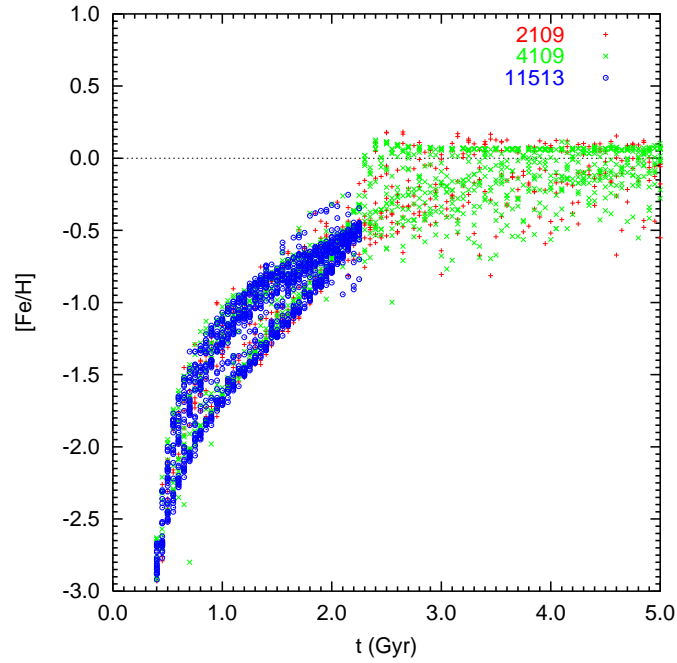}}
\caption{[Fe/H] evolution of the model galaxy with different "gas" particle number.}
\label{fe-t}
\end{figure}

\begin{figure}
\centerline{\includegraphics[width=3.5in]{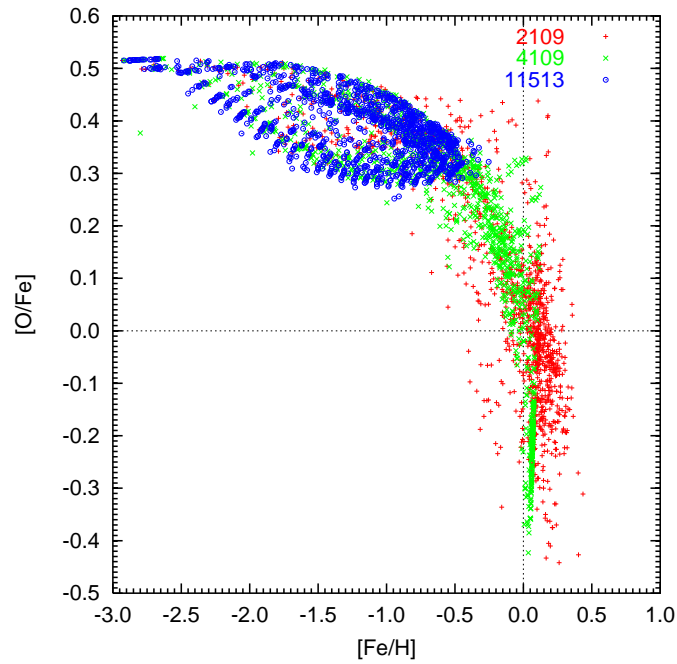}}
\caption{[O/Fe] vs. [Fe/H] of the model galaxy with different "gas" particle number.}
\label{ofe-fe}
\end{figure}

In Figure~\ref{fe-t} and Figure~\ref{ofe-fe} we present the chemical evolution
of our models. All three models have a very similar chemical history and
metallicity distribution.

\begin{figure}
\tabcapfont
\centerline{%
\begin{tabular}{c@{\hspace{0.1in}}c}
\includegraphics[width=2.5in]{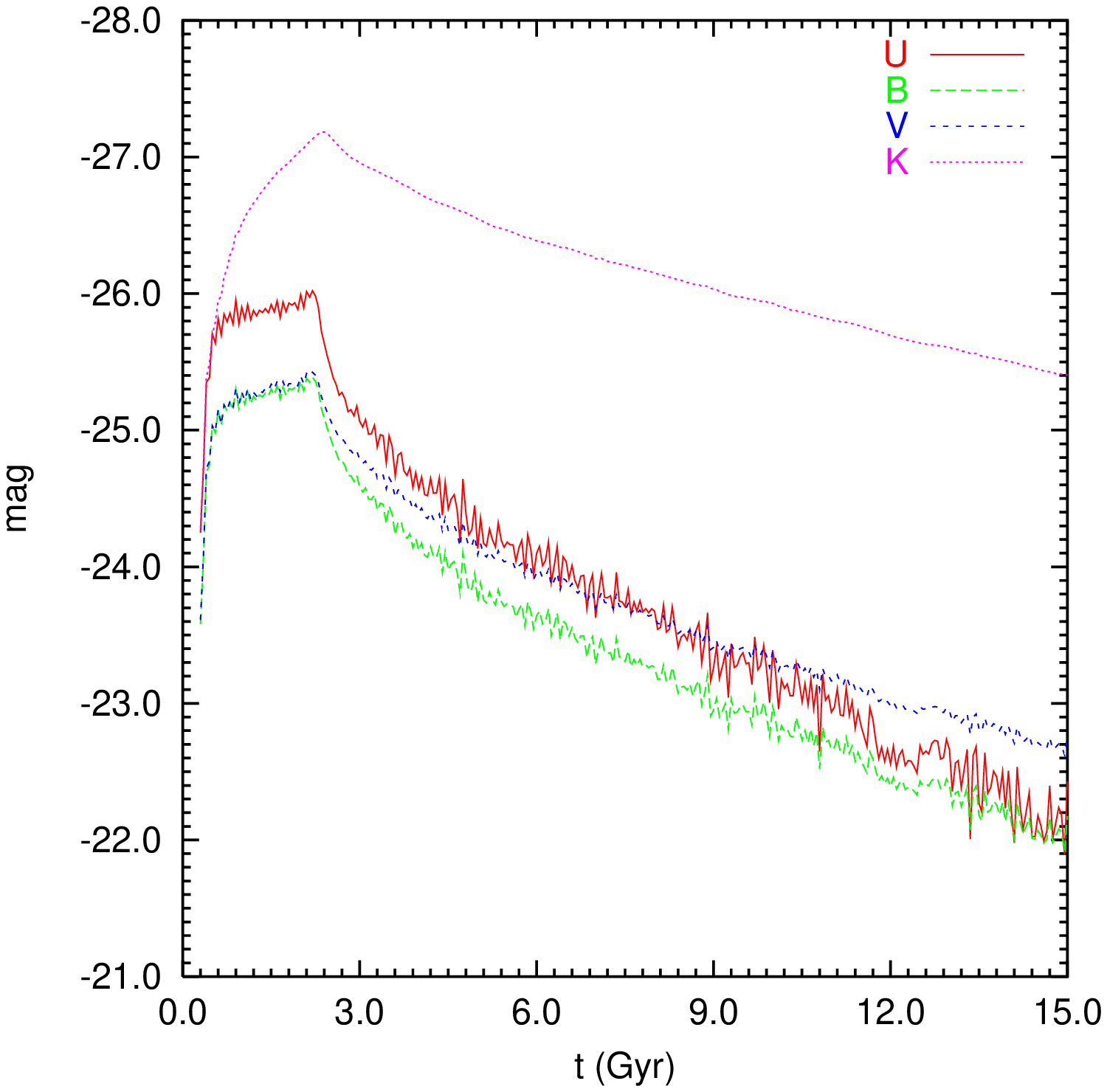} &
\includegraphics[width=2.5in]{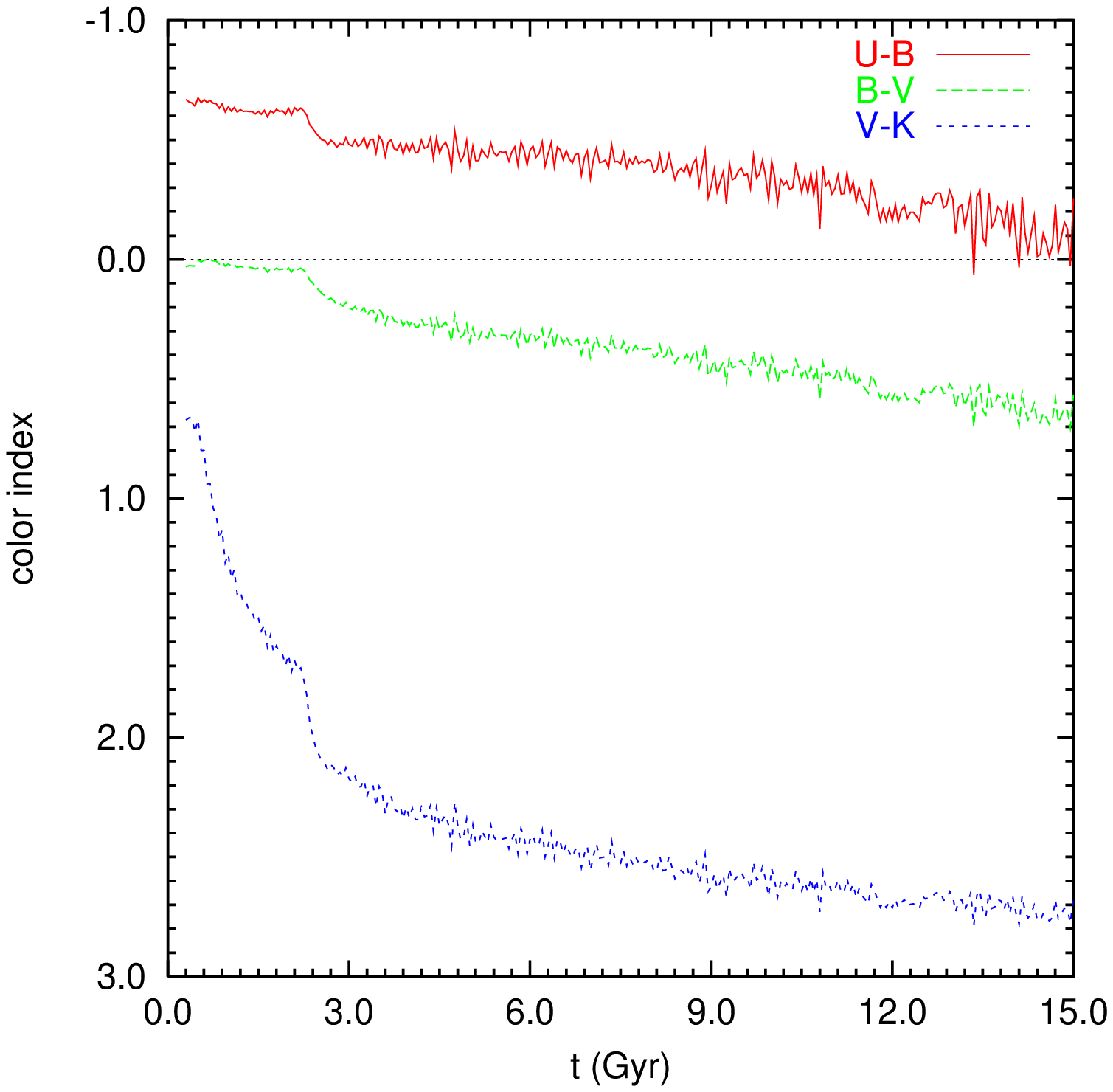} \\
a.~U, B, V, K evolution. & b.~(U-B), (B-V), (V-K) evolution.
\end{tabular}}
\caption{Photometric \& color evolution of the galaxy for the "basic" model.}
\label{mag&ci}
\end{figure}

The total photometric (in UBVK spectral brand) and color ((U-B), (B-V), (V-K))
evolution of our "base" model is presented in the Figure~\ref{mag&ci}a and
Figure~\ref{mag&ci}b. In case of our own Milky Way galaxy at the present day we
have a value for (B-V) $\approx 0.8$ \cite{vdK86}, what is close to our value
for model galaxy ($\approx 0.7$).

\begin{figure}
\tabcapfont
\centerline{%
\begin{tabular}{c@{\hspace{0.1in}}c}
\includegraphics[width=2.5in]{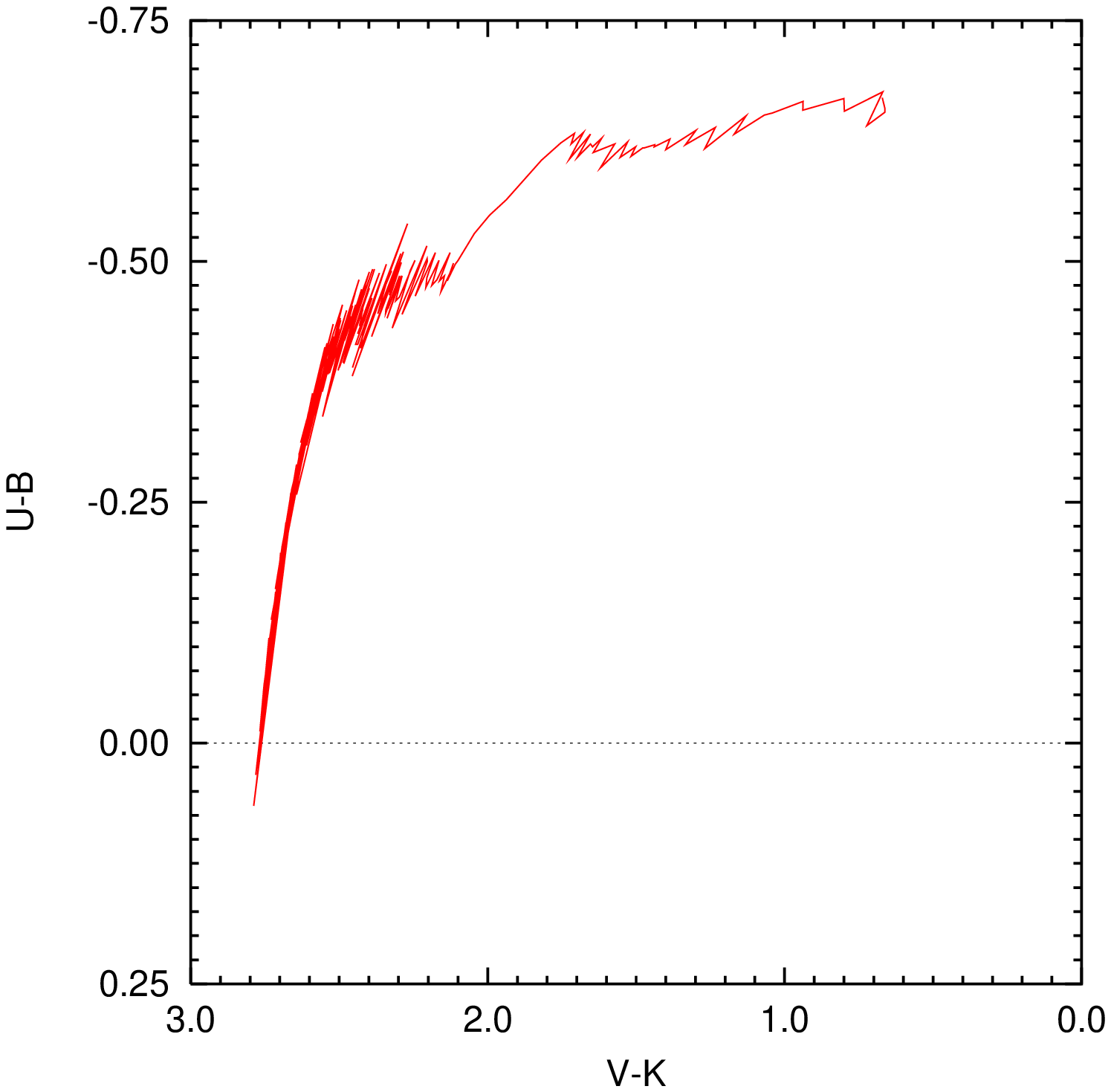} &
\includegraphics[width=2.5in]{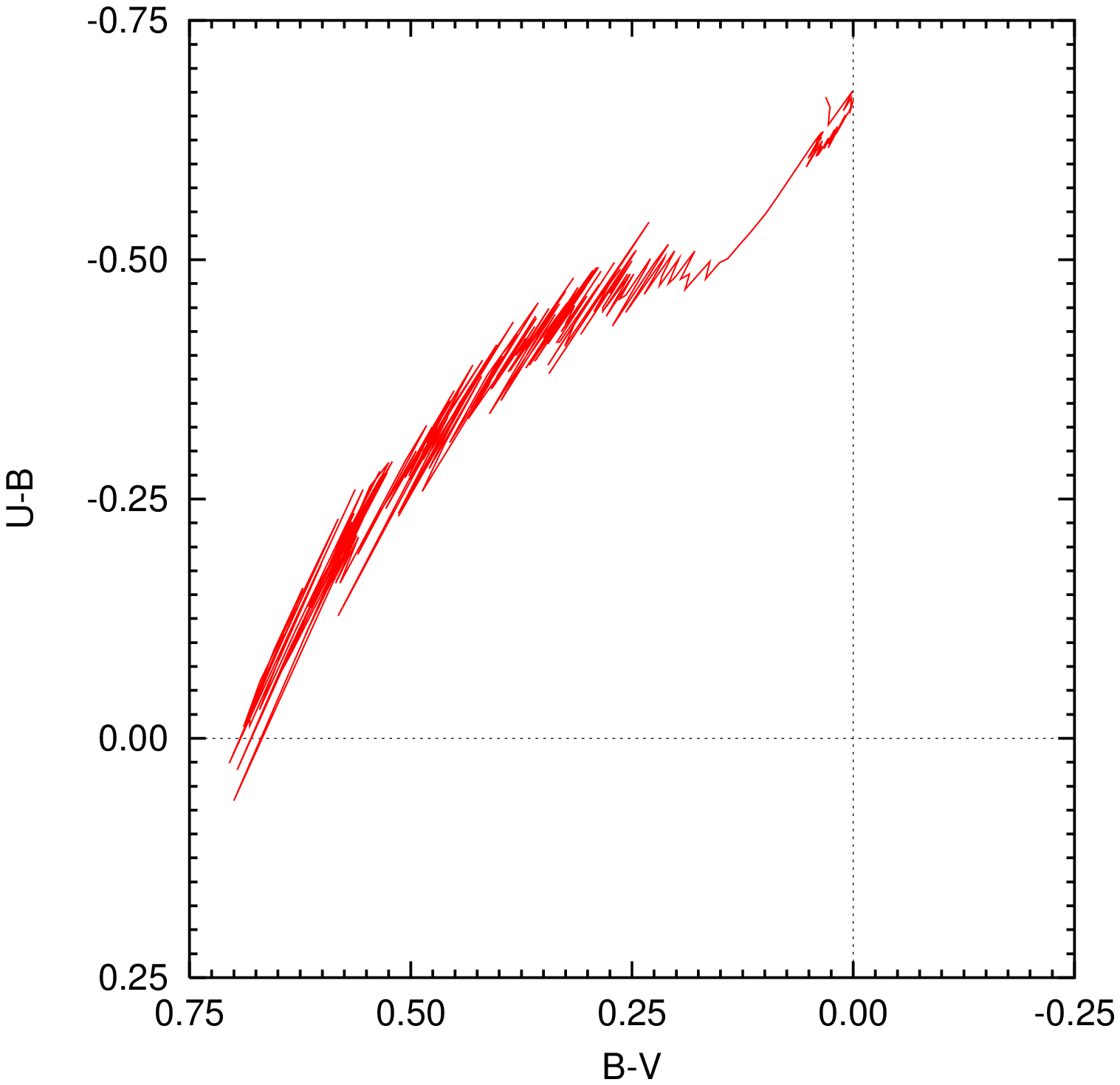} \\
a.~(U-B) vs. (V-K). & b.~(U-B) vs. (B-V).
\end{tabular}}
\caption{Color vs. color evolution of the galaxy for the "basic" model.}
\label{c&c}
\end{figure}

The color vs. color evolution of the model galaxy presented in
Figure~\ref{c&c}.


\section{Conclusion}

The presented model describes well the time evolution of the basic dynamical,
chemical and photometric parameters of a disk galaxy similar to the Milky Way.
The metallicity, luminosity and colors obtained are typical for such disk
galaxies. During the calculations we made an extended test of the proposed new
SF criteria. We find that the obtained results with different "gas" and "star"
particle numbers are not only qualitatively but also quantitatively similar.


\acknowledgements

The work was supported by the German Science Foundation (DFG) under grants
No. 436 UKR 18/2/99, 436 UKR 17/11/99 and partially supported by NATO
grant NIG 974675.

Special thanks for hospitality to the Astronomisches Rechen-Institut (ARI)
Heidelberg, where part of this work has been done.

It is a pleasure to thank Christian Theis for comments on an earlier version of
this paper.



\newpage

\newpage

\newpage

\newpage

\end{article}
\end{document}